\documentclass{article}
\begin{document}
\title{Modeling of diffusion processes on physical and structural levels 
       of materials}
\author{\Large A.~L.~Svistkov\thanks{E-mail: svistkov@icmm.ru}, 
        A.~V.~Ilinykh\\[1em]
        \itshape Institute of Continuous Media Mechanics,\\
        \itshape Academic Korolev Str.~1, 614013 Perm, Russia}
\date{}\maketitle
\begin{abstract}
A possibility to use an integral operator for establishing the link
between physical and structural levels of materials in modeling diffusion processes 
is considered. We show how to perform the transition 
from the stochastic description of motion of a system of points to 
the continuum description of diffusive profiles.
\end{abstract}

\section{Introduction}

Materials used industrially are complex multilevel systems. Analysis of these
systems generally involves the consideration of their physical, structural and
macroscopic levels (micro-, meso- and macrolevels). As a rule, for each level
we can formulate a special equation that allows modeling the processes taking place
in the material: molecular motions and interactions on the physical level,
supermolecular formations and processes in the vicinity of inclusions in
composites on the structural level, and the state of the body as a whole on
the macroscopic level.

The considered processes can be studied separately on each level, which
essentially simplifies their analysis. However, this possibility does not 
always exist.  Such a modeling can be realized in the case when 
the points of higher level are the material regions that are
representative (contain a great number of elements) and in homogeneous state. 
Under these conditions, the transition to state parameters of higher level 
can be performed by the averaging procedure. The studied volumes must be representative enough to
exclude the fluctuation of parameters. The homogeneity of the state of these
regions makes it possible to search the functional dependences on the
higher level.

There is another way of establishing the link between processes on different
levels. It suggests the construction of the mathematical operator
for creation of process images on the higher level instead of the use of the
hypothesis of representativeness and homogeneity of volumes. For composites,
the transition from the structural level to the macroscopic one by
the appropriate mathematical operator has been discussed in
\cite{Svistkov00a}, \cite{Svistkov00b}. This idea can also be used to realize 
the transition from the physical level to the structural one in modeling diffusion
phenomena. In the present paper, we consider the simplest case that leads to
the Fick law. The analysis of this case allows us to compare the results of
numerical modeling of diffusion of a great number of particles with the process
images on the level of continuous medium, to follow the disappearance of
fluctuations and to find the most suitable form of the mathematical operator.

\section{Modeling of diffusion of dissolved material constituents  
         on physical level}

We consider diffusion of $n$ material points (particles). Their motion in the
Euclidean space reproduces the diffusion of dissolved material constituents.
It is assumed that the change in the position of the material point
in space can be defined by Ito stochastic equations
\begin{equation}
  d{\bf x}_n\,=\,{\bf v}_n\,dt+\,b_n\,d{\bf z}_n,\hspace{2em}n=1,...N,
  \label{eq:hnjaw}
  \end{equation}
  $${\bf v}_n\,=\,{\bf v}_n({\bf x}_1,...,{\bf x}_N),\hspace{2em}
  b_n\,=\,b_n({\bf x}_1,...,{\bf x}_N),$$
where ${\bf x}_n$ is the radius-vector of the $n^{\rm th}$ particle at time $t$, 
${\bf v}_n$~is the vector function, which defines the determinate component of
particle motions in the Euclidean space, ${\bf z}_n$ are independent Wiener's
processes, and $b_n>0$ is the scalar function characterizing the peculiarities
of the chaotic motion of the $n^{\rm th}$ particle.

Let us clarify the obtained expressions. It is assumed, in the framework of
this model, that particles move chaotically and, as a result, redistribute
throughout the material, i.~e., the mass transfer takes place.
Constitutive equations are given as follows. The random motion can be
represented as the most probable tendency (velocity ${\bf v}_n$) and the
chance deviations from it. The vector ${\bf v}_n$ is a mathematical
expectation of the rate of change of the $n^{\rm th}$ particle space position. 
The deviations are described in the analysis of independent Wiener processes. The
scatter value for random walks is given by the function $b_n$.
Our purpose in this work is to determine, in the framework of used mathematical
models, conditions for the transition from the stochastic description
of motion of points to the contunuum description of mass transfer 
by the diffusion equation of a material constituent.

To derive the dependences we are interested in, the notion of a phase space
${\it\Gamma}$ should be used, which is a set of possible values of the
radius-vector components of particles. The phase space element
$d{\it\Gamma}$ is determined from
  $$d{\it\Gamma}\,=\,dx^1_1\,dx^2_1\,dx^3_1\,...\,
  dx^1_N\,dx^2_N\,dx^3_N\,,$$
where $x^i_n$~is the $i^{\rm th}$ coordinate of the radius-vector of $n^{\rm th}$ particle.
To every system state there corresponds the phase space point ${\it\Gamma}$.
From here on the symbol $\psi$ will be used to denote the probability density
of particles at time $t$ in phase space $\it\Gamma$ determined by the
radius-vectors ${\bf x}_1$,
..., ${\bf x}_N$   $$\psi\,=\,\psi(t,{\bf x}_1,...,{\bf x}_N).$$
The temporal variation of this density is defined by the Fokker~--- Plank
equation \cite{Gixman77}, \cite{Levi72}
  \begin{equation}
  \frac{\partial\psi}{\partial t}\,+\sum_{n=1}^N
  \nabla_{{\bf x}_n}\!\!\cdot\bigg(\psi\,{\bf v}_n\bigg)\,
  -\,\frac{1}{2}\sum_{n=1}^N\nabla^{\;2}_{{\bf x}_n}
  {\bf\,:\,}\bigg({b_n}^2\,{\bf I}\bigg)\,=0,
  \label{eq:FPK}
  \end{equation}
where
  $$\nabla_{{\bf x}_n}\,...\,=\,\sum^3_{i=1}
  {\bf i}_i\,\frac{\partial}{\partial x^i_n}\,...\,,\hspace{2em}
  \nabla^{\;2}_{{\bf x}_n}\,...\,=\,\nabla_{{\bf x}_n}
  \bigg(\nabla_{{\bf x}_n}\,...\,\bigg)\,,$$
${\bf I}$ is  a unit tensor, and ${\bf i}_i$ are the basis vectors of a rectangular
Cartesian coordinate system. It is clear that for phase space points 
at infinity
the probability density $\psi$ and its derivatives are equal to zero.

Numerical modeling of the motion of particles in space is limited to
prescribing the random walks of particles $\Delta{\bf x}_n$ within the time
interval $\Delta t$ according to the Gauss law $\psi_n$ with the probability
density \cite{Levi72} which has the form
  $$\psi_n\,=\,\prod^3_{i=1}\frac{1}{\sqrt{2\pi\Delta t\,}\,b_n\,}\,
  \exp\Bigg(-\,\frac{(\Delta x^i_n-v^i_n\Delta t)^2}
  {2{b_n}^2\Delta t}\Bigg)\,,$$
where symbols $v^i_n$ are the components of vectors ${\bf v}_n$ in a rectangular
Cartesian coordinate system.

\section{Fluctuating and determinate mass densities of diffusing material
         constituent}

In order to introduce continuum notions, we should perform 
the averaging procedure over the vicinity of the point in the Euclidean space. 
The simplest, yet not the best, way allowing the introduction of 
the mass density of a diffusing material constituent suggests the division of 
the number of particles in the vicinity of the studied point by the volume 
of this vicinity. This procedure gives the function dependent on 
the radius-vector of the Euclidean space, which pulsates as 
the radius-vector changes. The derivatives of this
function with respect to coordinates can not be determined in this case. The
region of averaging must be large enough to suppress these pulsations. The
transition to the smoothed differential continuum notions only be
possible following the corresponding hypothesis. In this section, we consider
more formalized way of inserting the mass density of a diffusing constituent in
our model.

Let us use the function $\it\Pi$ to introduce the mesolevel notions. By this
function one can evaluate the contribution of the $n^{\rm th}$ material point having
the radius-vector ${\bf x}_n$ at time $t$ to the state characteristics of
the continuous medium at the point in the 
Euclidean space having the radius-vector $\bf
x$. We require that the function $\it\Pi$ be continuous and have the
continuous first derivative and the piece-wise continuous second derivative
and become zero on moving apart from the point $\bf x$ to the
distance exceeding the value $a_r$
  \begin{equation}
  {\it\Pi}({\bf x}-{\bf x}_n)=0
  \hspace{1em}\mbox{при}\hspace{1em}
  \sqrt{({\bf x}-{\bf x}_n)\cdot\,({\bf x}-{\bf x}_n)}\ge a_r,
  \label{eq:dgdgdg}
  \end{equation}
satisfy the normalization condition
  \begin{equation}
  \int\limits^\infty_{-\infty}\!\int\limits^\infty_{-\infty}\!
  \int\limits^\infty_{-\infty}
  {\it\Pi}(x^1{\bf i}_1+x^2{\bf i}_2+x^3{\bf i}_3)
  \,dx^1\,dx^2\,dx^3=1,
  \label{eq:ooooo}
  \end{equation}
and the requirement of independence of the space orientation of basis vectors
  $${\it\Pi}(x^1{\bf i}_1+x^2{\bf i}_2+x^3{\bf i}_3)=
  {\it\Pi}\bigg({\bf Q}\cdot (x^1{\bf i}_1+x^2{\bf i}_2+
  x^3{\bf i}_3)\bigg),$$
where $x^i$ are the coordinates of a rectangular Cartesian coordinate system, and
and $\bf Q$ is the arbitrary rotation tensor. The function $\it\Pi$ must be
continuous and twice differentiated in order we can find the continuous twice
differentiated parameters of the medium. The equations must satisfy the
invariance requirement, i.~e., they must be independent of the orientation of
basis vectors. Hereafter, the function $\it\Pi$ of the argument
 ${\bf x}$--${\bf x}_n$ will be denoted by the symbol  ${\it\Pi}_n$
  $${\it\Pi}_n={\it\Pi}({\bf x}-{\bf x}_n).$$

Introduce the notion of the fluctuating density $\tilde\rho$ for the diffusing
material constituent by the following equality:
  $$\tilde\rho(t,{\bf x})\,=\,m\sum_{n=1}^N{\it\Pi}_n,$$
where $m$ is the mass of one diffusing particle. Physically, the calculation of
the density $\tilde\rho$ involves averaging the particles over the volume of
the material taking into account their distance from the considered point in
the Euclidean space. The value $\tilde\rho$ is the continuous and twice
differentiated function, as the function ${\it\Pi}$ is continuous and
twice differentiated.

Introduce the notion of the density $\rho$ of the diffusing constituent of the
material by the expression
  $$\rho\,=\,m\int\limits_{\it\Gamma}\sum_{n=1}^N
  {\it\Pi}_n\psi\,d{\it\Gamma}$$
which is a mathematical expectation of the fluctuating density.
Good agreement between $\tilde\rho$ and $\rho$ is only possible when the
parameter $a_r$ (\ref{eq:dgdgdg}) is sufficiently large and the distribution of
particles in the spherical vicinity of the radius $a_r$ of the considered point
in the Euclidean space is almost uniform.

\section{Derivation of the integral diffusion law from stochastic equations
         of particle motion}

Consider the mathematical expression
  \begin{equation}
  \int\limits_{\it\Gamma}\bigg(\sum^N_{k=1}{\it\Pi}_k\bigg)\,\Bigg[
  \frac{\partial\psi}{\partial t}+\sum_{n=1}^N\nabla_{{\bf x}_n}\!\!
  \cdot\bigg(\psi\,{\bf v}_n\bigg)\,-\frac{1}{2}\sum_{n=1}^N
  \nabla^{\;2}_{{\bf x}_n}{\bf\,:\,}
  \bigg({b_n}^2\psi\,{\bf I}\bigg)\Bigg]\,d{\it\Gamma}=0,
  \label{eq:hbtfrd}
  \end{equation}
Its validity follows from (\ref{eq:FPK}), as the integrand function (in
square brackets) is equal to zero. Let us rearrange equation (\ref{eq:hbtfrd})
using three identities. The first can be derived based on the time independence
of the function ${\it\Pi}_k$
  $$\int\limits_{\it\Gamma}\bigg(\sum^N_{k=1}{\it\Pi}_k\bigg)\,
  \frac{\partial\psi}{\partial t}\,d{\it\Gamma}\,=\,
  \int\limits_{\it\Gamma}\frac{\partial}{\partial t}
  \bigg(\sum^N_{k=1}{\it\Pi}_k\,\psi\bigg)\,d{\it\Gamma}.$$
The second is a consequence of vanishing of the probability density of state
$\psi$ at infinity of the phase space $\it\Gamma$.
  \begin{eqnarray}
  &&\hspace{-2em}
  \int\limits_{\it\Gamma}\bigg(\sum^N_{k=1}{\it\Pi}_k\bigg)\,
  \nabla_{{\bf x}_n}\!\!\cdot\bigg(\psi\,{\bf v}_n\bigg)\,d{\it\Gamma}\,=
  \nonumber\\&&\nonumber\\
  &=&\int\limits_{\it\Gamma}\nabla_{{\bf x}_n}\!\!\cdot
  \Bigg[\bigg(\sum^N_{k=1}{\it\Pi}_k\bigg)\,
  \bigg(\psi\,{\bf v}_n\bigg)\Bigg]\,d{\it\Gamma}\,-
  \int\limits_{\it\Gamma}\nabla_{{\bf x}_n}
  \bigg(\sum^N_{k=1}{\it\Pi}_k\bigg)\cdot
  \bigg(\psi\,{\bf v}_n\bigg)\,d{\it\Gamma}\,=
  \nonumber\\&&\nonumber\\
  &=&-\int\limits_{\it\Gamma}\nabla_{{\bf x}_n}
  \bigg(\sum^N_{k=1}{\it\Pi}_k\bigg)\cdot
  \bigg(\psi\,{\bf v}_n\bigg)\,d{\it\Gamma}\,.
  \nonumber\end{eqnarray}
The last is derived from the expression
  \begin{eqnarray}
  &&\hspace{-2em}
  \int\limits_{\it\Gamma}\bigg(\sum^N_{k=1}{\it\Pi}_k\bigg)\,
  \nabla^{\;2}_{{\bf x}_n}
  {\bf\,:\,}\bigg({b_n}^2\psi\,{\bf I}\bigg)\,d{\it\Gamma}\,=
  \nonumber\\&&\nonumber\\
  &=&\int\limits_{\it\Gamma}\nabla_{{\bf x}_n}\!\!\cdot
  \Bigg[\bigg(\sum^N_{k=1}{\it\Pi}_k\bigg)\,\nabla_{{\bf x}_n}\!\!\cdot
  \bigg({b_n}^2\psi\,{\bf I}\bigg)\Bigg]\,d{\it\Gamma}\,-
  \nonumber\\&&\nonumber\\
  &&-\int\limits_{\it\Gamma}\nabla_{{\bf x}_n}
  \bigg(\sum^N_{k=1}{\it\Pi}_k\bigg)\cdot\Bigg[
  \nabla_{{\bf x}_n}\!\!\cdot
  \bigg({b_n}^2\psi\,{\bf I}\bigg)\Bigg]\,d{\it\Gamma}\,=
  \nonumber\\&&\nonumber\\
  &=&-\int\limits_{\it\Gamma}\nabla_{{\bf x}_n}
  \bigg(\sum^N_{k=1}{\it\Pi}_k\bigg)\cdot\Bigg[
  \nabla_{{\bf x}_n}\!\!\cdot\bigg({b_n}^2\psi\,{\bf I}\bigg)
  \Bigg]\,d{\it\Gamma}
  \nonumber\end{eqnarray}
and has the form
  \begin{eqnarray}
  &&\hspace{-2em}
  \int\limits_{\it\Gamma}\bigg(\sum^N_{k=1}{\it\Pi}_k\bigg)\,
  \nabla^{\;2}_{{\bf x}_n}
  {\bf\,:\,}\bigg({b_n}^2\psi\,{\bf I}\bigg)\,d{\it\Gamma}\,=
  \nonumber\\&&\nonumber\\
  &=&-\int\limits_{\it\Gamma}\nabla_{{\bf x}_n}\!\!\cdot
  \Bigg[\nabla_{{\bf x}_n}
  \bigg(\sum^N_{k=1}{\it\Pi}_k\bigg)\cdot
  \bigg({b_n}^2\psi\,{\bf I}\bigg)\Bigg]\,d{\it\Gamma}\,+
  \nonumber\\&&\nonumber\\
  &&+\int\limits_{\it\Gamma}\nabla^{\;2}_{{\bf x}_n}
  \bigg(\sum^N_{k=1}{\it\Pi}_k\bigg){\bf\,:\,}
  \bigg({b_n}^2\psi\,{\bf I}\bigg)\,d{\it\Gamma}\,=
  \nonumber\\&&\nonumber\\
  &=&\int\limits_{\it\Gamma}\nabla^{\;2}_{{\bf x}_n}
  \bigg(\sum^N_{k=1}{\it\Pi}_k\bigg){\bf\,:\,}
  \bigg({b_n}^2\psi\,{\bf I}\bigg)\,d{\it\Gamma}\,.
  \nonumber\end{eqnarray}
With these identities, equality (\ref{eq:hbtfrd}) becomes
  \begin{eqnarray}
  &&\hspace{-2em}
  \int\limits_{\it\Gamma}\frac{\partial}{\partial t}
  \Bigg[\bigg(\sum^N_{k=1}{\it\Pi}_k\bigg)\,\psi
  \Bigg]\,d{\it\Gamma}-\int\limits_{\it\Gamma}
  \sum^N_{n=1}\nabla_{{\bf x}_n}
  \bigg(\sum^N_{k=1}{\it\Pi}_k\bigg)\cdot
  \bigg(\psi\,{\bf v}_n\bigg)\,d{\it\Gamma}\,-
  \nonumber\\&&\nonumber\\
  &&-\frac{1}{2}\,\int\limits_{\it\Gamma}\sum^N_{n=1}
  \nabla^{\;2}_{{\bf x}_n}\bigg(\sum^N_{k=1}{\it\Pi}_k\bigg){\bf\,:\,}
  \bigg({b_n}^2\psi\,{\bf I}\bigg)\,d{\it\Gamma}\,=0.
  \nonumber\end{eqnarray}
The replacement of the operator $\nabla_{{\bf x}_n}$ by the operator
$-\nabla$ for the function $\it\Pi$
  $$\nabla_{{\bf x}_n}\bigg(\sum^N_{k=1}{\it\Pi}_k\bigg)\,=\,
  \nabla_{{\bf x}_n}{\it\Pi}_n\,=\,
  -\,\nabla{\it\Pi}_n,$$
gives
  \begin{eqnarray}
  &&\hspace{-2em}
  \int\limits_{\it\Gamma}\frac{\partial}{\partial t}
  \Bigg[\bigg(\sum^N_{n=1}{\it\Pi}_n\bigg)\,\psi
  \Bigg]\,d{\it\Gamma}+\int\limits_{\it\Gamma}
  \sum^N_{k=1}\nabla{\it\Pi}_n\cdot
  \bigg(\psi\,{\bf v}_n\bigg)\,d{\it\Gamma}\,-
  \nonumber\\&&\nonumber\\
  &&-\frac{1}{2}\,\int\limits_{\it\Gamma}\sum^N_{n=1}
  \nabla^2{\it\Pi}_n{\bf\,:\,}
  \bigg({b_n}^2\psi\,{\bf I}\bigg)\,d{\it\Gamma}\,=0.
  \nonumber\end{eqnarray}
where
  $$\nabla\,...\,=\,\sum^3_{i=1}
  {\bf i}_i\,\frac{\partial}{\partial x^i}\,...\,,\hspace{2em}
  \nabla^2\,...\,=\,\nabla
  \bigg(\nabla\,...\,\bigg)\,.$$

Taking the operations $\partial/\partial t$ and $\nabla$ out of the integration
sign, we obtain the expression in question
  \begin{eqnarray}
  &&\hspace{-2em}
  \frac{\partial}{\partial t}\int\limits_{\it\Gamma}
  \sum^N_{n=1}{\it\Pi}_n\,\psi\,d{\it\Gamma}+
  \nabla\cdot\int\limits_{\it\Gamma}
  \sum^N_{n=1}{\it\Pi}_n\,\psi\,{\bf v}_n\,d{\it\Gamma}\,-
  \nonumber\\&&\nonumber\\
  &&-\frac{1}{2}\,\nabla^2{\bf\,:\,}
  \int\limits_{\it\Gamma}\sum^N_{n=1}{\it\Pi}_n\,
  {b_n}^2\psi\,{\bf I}\,d{\it\Gamma}\,=0.
  \label{eq:ttffcx}
  \end{eqnarray}
It can be used to derive the sufficient condition for transition to modeling
the mass transfer based on the diffusion equation of the medium constituent.

\section{Sufficient conditions for transition to continuum notions}

The diffusion process can not be modeled using the notion of the chemical
potential in the framework of the proposed approach. Such modeling must be based
on the analysis of the first and second thermodynamic laws, the accounting of
the interaction forces between the particles in the formulation of constitutive
equations, and so on. With stochastic equations (\ref{eq:hnjaw}), the transition
is possible to the Fick diffusion law only.

Define the notion of a diffusion coefficient $D$ by the equality
  $$\rho\,D\,=\,\frac{m}{2}\,\int\limits_{\it\Gamma}
  \sum^N_{n=1}{\it\Pi}_n\,{b_n}^2\psi\,d{\it\Gamma}.$$
To describe the process on the meso-level, it is sufficient to satisfy two
conditions: the vectors ${\bf v}_n$ must obey the equality
  \begin{equation}
  \rho\,{\bf v}\,+\,\rho\,\nabla D\,=\,m\,
  \int\limits_{\it\Gamma}\sum^N_{n=1}
  {\it\Pi}_n\,{\bf v}_n\,\psi\,d{\it\Gamma},
  \label{eq:kmjnub}
  \end{equation}
where $\bf v$ is the center-of-mass velocity of the body and the functional
relation must exist between two values $\rho$ and $D$
  \begin{equation}
  D\,=\,D(\rho).
  \label{eq:kmjnuba}
  \end{equation}
With these requirements satisfied, equation (\ref{eq:ttffcx}) can be written
in the form
  $$\frac{\partial\rho}{\partial t}\,+\,
  \nabla\cdot(\rho\,{\bf v})\,=\,\nabla\cdot
  \bigg(D\,\nabla\rho\bigg)\,.$$
which is a well-known Fick law. The possibility to describe the process by the
equation of continuous medium is checked, in particular problems, by conditions
(\ref{eq:kmjnub}) and (\ref{eq:kmjnuba}). It should be noted that we are
dealing not with the averaged values, but with their mathematical expectations.
In practice it means that these values should be calculated by averaging over
space and random realizations.

\section{Integral operator kernel of third order accuracy of image 
         reproducing}

For problem solution we should define a particular form of the integral 
operator kernel ${\it\Pi}$.
We take the distance $r$ between the radius-vectors $\bf x$ and ${\bf x}_*$ 
as an argument of function ${\it\Pi}$
  $${\it\Pi}\,=\,{\it\Pi}(r),\hspace{2em}
  r\,=\,\sqrt{({\bf x}-{\bf x}_*)\cdot\,({\bf x}-{\bf x}_*)}.$$
To reproduce the effective density of particles most precisely, it is
reasonable to use the function ${\it\Pi}(r)$ based on the ideas of
constructing the integral operator kernel described by the RKPM method
\cite{Chen96}, \cite{Chen97}. 

The kernel ${\it\Pi}$ is supposed to meet the following requirements.
Let some function $\beta({\bf x})$ be represented as the coordinate 
dependence of the third order and the fast oscillating term
$\gamma({\bf x})$
  $$\beta({\bf x})\,=\,b+{\bf b}\cdot{\bf x}+{\bf B}{\bf\,:\,}
  {\bf x}\otimes{\bf x}+\stackrel{3}{\bf B}\vdots\,{\bf x}
  \otimes{\bf x}\otimes{\bf x}+\gamma(x),$$
where $b$, $\bf b$, $\bf B$, $\stackrel{3}{\bf B}$ are tensors of zero, first, 
second and third rank. The cubic dependence must remain unchanged for any 
space point ${\bf x}_*$ under the transition to the image performed by the 
integral operator
  \begin{eqnarray}
  &&\int\limits_V{\it\Pi}(r)\,\bigg(\,b+{\bf b}\cdot{\bf x}+
  {\bf B}{\bf\,:\,}{\bf x}\otimes{\bf x}+
  \stackrel{3}{\bf B}\vdots\,{\bf x}\otimes{\bf x}\otimes{\bf x}
  \bigg)\,dV\,=
  \nonumber\\&&\hspace{2em}
  =\,b+{\bf b}\cdot{\bf x}_*+{\bf B}{\bf\,:\,}{\bf x}_*\otimes{\bf x}_*+
  \stackrel{3}{\bf B}\vdots\,{\bf x}_*\otimes{\bf x}_*\otimes{\bf x}_*.
  \label{eq:qwe}
  \end{eqnarray}
The symbol $V$ stands for the entire volume of the Euclidean space. 
In the problems where the fast pulsations of the function 
$\gamma({\bf x})$ suppress each other under integral transformation, i.~e,
  $$\int\limits_V{\it\Pi}(r)\,\gamma({\bf x})\, dV\,\approx 0,$$
the smoothed integral image of the initial function 
$\beta({\bf x})$ will be the slowly changing cubic function 
  $$\int\limits_V{\it\Pi}(r)\,\beta({\bf x})\,dV\,\approx
  \,b+{\bf b}\cdot{\bf x}_*+{\bf B}{\bf\,:\,}{\bf x}_*\otimes{\bf x}_*+
  \stackrel{3}{\bf B}\vdots\,{\bf x}_*\otimes{\bf x}_*\otimes{\bf x}_*.$$
In this case the integral operator will work as a "filter" that keeps 
slowly changing regularities and cancels fast pulsations.

Determine the conditions which the integral operator of the third order
accuracy should satisfy. The function $\beta({\bf x})$ may be developed  
as a series in powers of $\Delta{\bf x}$ about point ${\bf x}$
  \begin{eqnarray}
  \beta({\bf x})&=&b+{\bf b}\cdot{\bf x}+
  {\bf B}{\bf\,:\,}{\bf x}\otimes{\bf x}+\stackrel{3}{\bf B}\vdots\,
  {\bf x}\otimes{\bf x}\otimes{\bf x}+
  \nonumber\\&&
  +\,a+{\bf a}\cdot\Delta{\bf x}+{\bf A}{\bf\,:\,}\Delta{\bf x}\otimes\Delta{\bf x} 
  +\stackrel{3}{\bf A}\vdots\,\Delta{\bf x}\otimes\Delta{\bf x}\otimes 
  \Delta{\bf x}+\gamma({\bf x}),
  \nonumber\end{eqnarray}
where 
  $$\Delta{\bf x}\,=\,{\bf x}-{\bf x}_*\,=\,
  \Delta x^1{\bf i}_1+\Delta x^2{\bf i}_2+\Delta x^3{\bf i}_3,$$
$a$, $\bf a$, $\bf A$, $\stackrel{3}{\bf A}$ are tensors of zero, first, 
second and third rank. We note that equality (\ref{eq:qwe}) will be true  
only when the reproducing conditions are fulfilled
  \begin{equation}
  \int\limits_V{\it\Pi}(r)dV=1,\hspace{2em}
  \int\limits_V{\it\Pi}(r)\,\Delta{\bf x}\,dV =0,
  \label{eq:llllkk}
  \end{equation}
  \begin{equation}
  \int\limits_V{\it\Pi}(r)\,\Delta{\bf x}\otimes\Delta{\bf x}\,dV\,=0,
  \hspace{2em}\int\limits_V{\it\Pi}(r)\,\Delta{\bf x}\otimes\Delta{\bf x}\, 
  \otimes\Delta{\bf x}\,dV\,=0.
  \label{eq:llllkkk}
  \end{equation}

It is easy to find that in the spherical system of coordinates 
\begin{eqnarray}
  \Delta x^1&=&r\cos(\alpha)\sin(\theta),
  \nonumber\\
  \Delta x^2&=&r\sin(\alpha)\sin(\theta),
  \nonumber\\
  \Delta x^3&=&r\cos(\theta) 
  \nonumber\end{eqnarray}
the reproducing conditions (\ref{eq:llllkk}), (\ref{eq:llllkkk}) are
equivalent to the requirement that the following equalities are valid
  $$\int\limits_0^{a_r}r^2\,{\it\Pi}(r)\,dr\,=\,\frac{1}{4\pi}\,, 
  \hspace{2em}
  \int\limits_0^{a_r}r^4\,{\it\Pi}(r)\,dr\,=0.$$
An example of the operator kernel of third order accuracy is the
function
  $${\it\Pi}(r)\,=\,\frac{1}{a_r^3}\,\bigg(\gamma_0+\gamma_1\,
  \frac{r^2}{a_r^2}\bigg)\bigg(1-\frac{r^2}{a_r^2}\bigg)^2\,H(a_r-r)$$
in which the constants $\gamma_0$ и $\gamma_1$ are chosen such that 
equations (\ref{eq:llllkk}), (\ref{eq:llllkkk}) are satisfied and the
symbol $H(...)$ denotes the Heavyside function.

\section{Integral operator kernel of first order accuracy of image 
            reproducing}

The transition from physical to structural level may be performed using
integral operators of first order accuracy. These operators must meet
the requirement that only linear dependences retain in integral
transformation
  $$\int\limits_V{\it\Pi}(r)\,\bigg(\,b+{\bf b}\cdot{\bf x}\bigg)\,dV\,=
  \,b+{\bf b}\cdot{\bf x}_*.$$
An example of integral operator kernels of first order accuracy is the
function
  $${\it\Pi}(r)\,=\,\frac{\gamma}{a_r^3}\,
  \bigg(1-\frac{r^2}{a_r^2}\bigg)^2\,H(a_r-r)$$
and
  \begin{equation}
  {\it\Pi}=\,\left\{
  \begin{array}{ll}
  \frac{\mbox{$\vphantom{p_p}\gamma$}}{\mbox{$\vphantom{A^A}a_r^3$}}\,,
  &\hspace{2em}0\le r<a_\gamma;\\&\\
  \frac{\mbox{$\vphantom{p_p}\gamma$}}{\mbox{$\vphantom{A^A}2a_r^3$}}\,
  \Bigg(1+\cos\bigg(\pi\frac{\mbox{$r-a_\gamma$}}
  {\mbox{$a_r-a_\gamma$}}\bigg)\Bigg)\,,
  &\hspace{2em}a_\gamma\le r<a_r;\\&\\
  0,&\hspace{2em}a_r\le r,\\
  \end{array}
  \right.
  \label{eq:hbvfrds}
  \end{equation}
in which constants $\gamma$ are determined from the normalization condition
  $$\int\limits_0^{a_r}r^2\,{\it\Pi}(r)\,dr\,=\,\frac{1}{4\pi}\,.$$

\section{Numerical experiments}

In this section, we consider the problem of penetration of diffusing particles
into a semi-infinite material with constant concentration of particles on the
material boundary. The constant $b$ will be used as a function $b_n$
  $$b_n\,=\,b,\hspace{2em}n=1,...,N,$$
and all vectors $\bf v$ and ${\bf v}_n$ are assumed to be zero
  $${\bf v}\,=\,0,\hspace{2em}
  {\bf v}_n\,=\,0,\hspace{2em}n=1,...,N.$$
In this case, equations (\ref{eq:kmjnub}) and (\ref{eq:kmjnuba}) are valid. The
diffusion coefficient is found by the formula
  $$D\,=\,\frac{{b}^2}{2}\,.$$

\begin{figure}
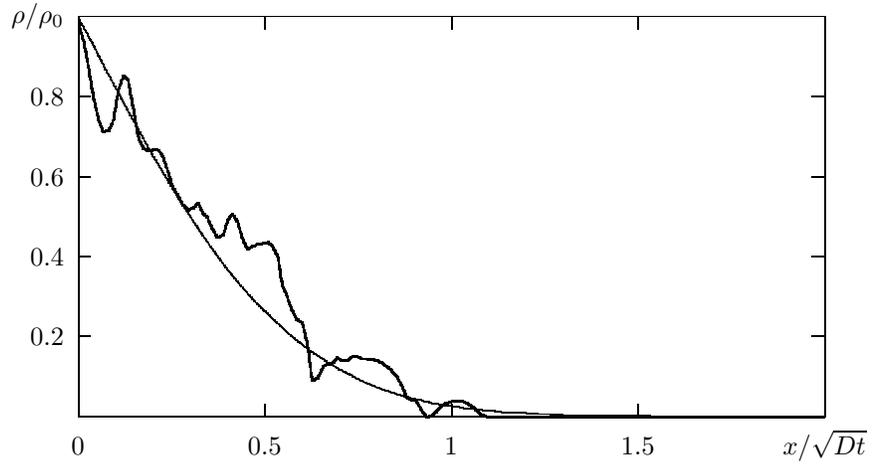

\noindent\mbox{}\hspace{-4em}
\input fig_1a.tex\\\mbox{}\hfil a\hfil\\
\noindent\mbox{}\hspace{-4em}
\input fig_1b.tex\\\mbox{}\hfil b\hfil
\caption{\label{aaa}
Comparison of the theoretical diffusive profile with the numerical results
when the concentration of diffusing particles on the material boundary is (a)
5000 per unit volume and (b) 30 per unit volume}  
\end{figure}

\begin{figure}
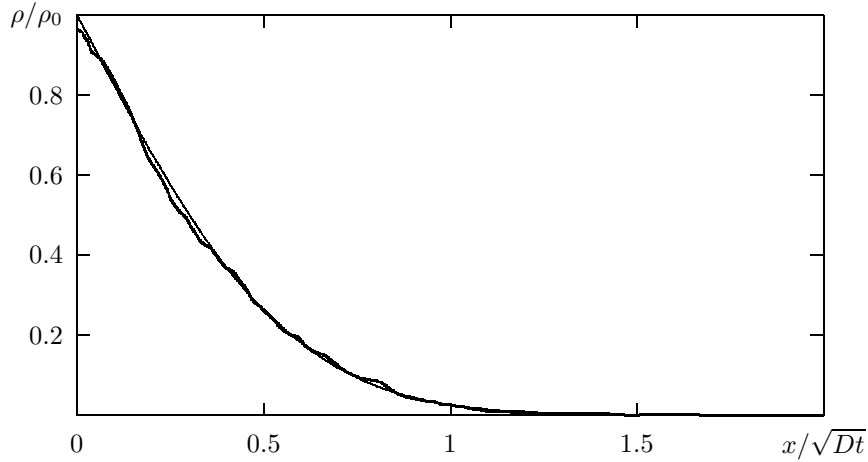

\noindent\mbox{}\hspace{-4em}
\input fig_2.tex
\caption{\label{ccc}
Comparison of the theoretical diffusive profile with the numerical results when
the concentration of diffusing particles on the material boundary is 30 per unit
volume. The results were averaged over 30 random realizations}
\end{figure}

The performed numerical experiments demonstrated that the best smoothing effect 
can be achieved with the aid of the function ${\it\Pi}(r)$ of first order accuracy of image 
reproducing (\ref{eq:hbvfrds}). This function must be constant in the region 
occupying the half of the spherical volume of radius $a_r$. In the present 
work, graphs (Fig.~\ref{aaa}, Fig.~\ref{ccc}) have been plotted based on the 
relation (\ref{eq:hbvfrds}).

The obtained numerical results led us to the conclusion that the analytical 
solution of the Fick equation \cite{Vladimirov81} is in good agreement with the fluctuating density
$\tilde\rho$ only in the case of a great number of particles
in the averaging region. For example, about 5000 particles entered the
radius sphere $a_r$ on the material boundary when we calculated the
dependence shown in Fig.~\ref{aaa}.a.


An essential divergence between the fluctuating density  $\tilde\rho$ and the
analytical solution can be observed under random realization of the process,
in the course of which 30 particles enter the sphere with radius
$a_r$ on the material boundary (Fig.~\ref{aaa}.b). However, one can also
obtain the mesolevel images of the process. For this, it is sufficient to
perform the averaging of the fluctuating density over thirty random
realizations. The obtained density $\left<\rho\right>$ agrees fairly well
with the analytical solution of the Fick equation (Fig.~\ref{ccc}) and is
close to its most probable value (mathematical expectation, i.~e. the
density $\rho$).

\section{Conclusion}

The state parameters for the medium studied (mass density, diffusion coefficients) can be
calculated with the aid of operators acting on a multitude of microscopic
parameters $a_1$, ..., $a_N$ and having the form
  $$L_1(a_1,...,a_N)\,=\,\sum^N_{n=1}{\it\Pi}_n\,a_n.$$
and
  $$L_2(a_1,...,a_N)\,=\,\int\limits_{\it\Gamma}
  \sum^N_{n=1}{\it\Pi}_n\,a_n\,\psi\,d{\it\Gamma},$$
where $a_n$ is the characteristic of the $n^{\rm th}$ paticle. 
The fist operator represents the averaging procedure in the vicinity of
the considered point in the Euclidean space and leads to the fluctuating
twice differentiated variables. The second operator gives the determinate
continuous twice-differentiated parameters, which can be used as the state
characteristics of a continuous medium. Determination of diffusive profiles
analogous to those on the mesolevel based on the fluctuating mass density
is possible provided that thousands of particles enter the region of
averaging. However, there is another way of constructing the mesolevel
images of the process. It consists in averaging over the fluctuating
densities calculated in numerical experiments. The results achieved in this
case agree fairly well with the calculation data obtained on the level of
continuous medium with much smaller number of particles entering the
averaging region.
\bigskip

The work has been done under the FPP "Integration" (project 00-01, grant
contest for students and young researchers of Perm region).



\begin{thebibliography}{99}

\bibitem{Svistkov00a} Svistkov A. L. Boundary conditions on internal
surfaces in a particulate elastomeric material // Proceedings XXVIII
Summer School, St. Petersburg (Repino), 2000 (to be published).

\bibitem{Svistkov00b} Svistkov A. L., Komar L. A., Ilinykh A. V. Integral
operator for generation of macroscopic images of microstructural composite
parameters // Proceedings XXVIII Summer School, St. Petersburg (Repino),
2000 (to be published).

\bibitem{Gixman77} Gikhman~I.~I., Skorohod~A.~V. Introduction to the theory
of random processes.~--- Moscow: Nauka. 1977.~--- 568~p. (in Russian).

\bibitem{Levi72} Levi~P. Stochastic processes and Brownian motion.
Moscow: Nauka. 1972.~--- 376~p. (in Russian).

\bibitem{Chen96} Chen~J.-S., Pan~C., Wu~C.-T., Liu~W.~K. Reproducing
kernel particle methods for large deformation analysis of non-linear
structures~// Comput. Methods Appl. Mech. Engrg.~--- 1996.~---
V.~139.~--- P.~195--227.

\bibitem{Chen97} Chen~J.-S., Pan~C., Wu~C.-T. Large deformation analysis
of rubber based on a reproducing kernel particle method~// Comput.
Mech.~--- 1997.~--- V.~19.~--- P.~211--227.

\bibitem{Vladimirov81} Vladimirov~V.~S. Equations from mathematical physics~---
Moscow: Nauka. 1981.~--- 512~p. (in Russian).

\end{thebibliography}
\end{document}